\documentclass[sigconf]{acmart}
\pdfoutput=1
\usepackage[utf8]{inputenc}

\usepackage{algorithmic}
\usepackage{cleveref}
\usepackage{comment}
\usepackage{csvsimple}
\usepackage{lipsum}
\usepackage{listings}
\usepackage{graphicx}
\usepackage{siunitx}
\usepackage{subfigure}
\usepackage{textcomp}
\usepackage{todonotes}
\usepackage{xcolor}
\usepackage{xspace}

\theoremstyle{definition}

\crefname{researchquestion}{RQ}{RQs}

\newcommand{\SWH}{Software Heritage\xspace}

\def\DataBlobCountApproxM{6.5}
\def\DataBlobCount{6482295}

\def\DataBlobNoOriginPct{12}

\def\DataBlobNoEarliestPct{11}
\def\DataTarSizeApprox{14\,GiB}
\def\DataMimeTextPlainPct{84}
\def\DataMimeTextPct{98}

 \copyrightyear{2022}
\acmYear{2022}
\setcopyright{acmlicensed}
\acmConference[MSR '22]{19th International Conference on Mining Software Repositories}{May 23--24, 2022}{Pittsburgh, PA, USA}
\acmBooktitle{19th International Conference on Mining Software Repositories (MSR '22), May 23--24, 2022, Pittsburgh, PA, USA}
\acmPrice{15.00}
\acmDOI{10.1145/3524842.3528491}
\acmISBN{978-1-4503-9303-4/22/05}

\title{A Large-scale Dataset of (Open Source) License Text Variants}
\author{Stefano Zacchiroli}
\email{stefano.zacchiroli@telecom-paris.fr}
\orcid{0000-0002-4576-136X}
\affiliation{\institution{LTCI, Télécom Paris, Institut Polytechnique de Paris}
  \city{Paris}
  \country{France}
}

\begin{abstract}

  We introduce a large-scale dataset of the complete texts of free/open source
  software (FOSS) license variants. To assemble it we have collected from the
  Software Heritage archive---the largest publicly available archive of FOSS
  source code with accompanying development history---all versions of files
  whose names are commonly used to convey licensing terms to software users and
  developers.

  The dataset consists of \DataBlobCountApproxM{} million unique license files
  that can be used to conduct empirical studies on open source licensing,
  training of automated license classifiers, natural language processing (NLP)
  analyses of legal texts, as well as historical and phylogenetic studies on
  FOSS licensing.

  Additional metadata about shipped license files are also provided, making the
  dataset ready to use in various contexts; they include: file length measures,
  detected MIME type, detected SPDX license (using ScanCode), example origin
  (e.g., GitHub repository), oldest public commit in which the license
  appeared.

  The dataset is released as open data as an archive file containing all
  deduplicated license files, plus several portable CSV files for metadata,
  referencing files via cryptographic checksums.

\end{abstract}

\keywords{dataset, open source, software license, copyright, intellectual
  property, software engineering, natural language processing}

\begin{document}
\maketitle

\section{Introduction}
\label{sec:intro}

Free/Open Source Software (FOSS) is ubiquitous in modern IT
solutions~\cite{synopsis2020ossra}. Its liberal licensing terms allow everyone,
including industry players, to reuse, build open, and extend it, subject to
conditions that vary from license to license~\cite{rosen2005osslicensing,
  lindberg2008osslicensing}.

Many different software licenses exist and are used in public code. Some of
those licenses are labeled as proper ``open source'' by the Open Source
Initiative, others (with a significant overlap) as ``free software'' by the
Free Software Foundation, others are neither ``open'' nor ``free'' but are
applied to software components distributed in source code form (e.g., via
GitHub or GitLab) and need to be dealt with when reusing those components. The
ecosystem of licensing terms is so varied that industry standards like SPDX
emerged to normalize license naming and identifiers~\cite{stewart2010spdxspec}.

Proper management of such an increasingly complex software supply
chain~\cite{harutyunyan2020osssupplychain} requires being able to deal with
license combinations, their potential
incompatibility~\cite{german2012liccompatibility}, and auditing increasingly
large code bases, ideally in an automated
way~\cite{phipps2020continuouscompliance}.

These real-world needs have motivated over the years several empirical software
engineering (ESE) studies on the evolution of open source
licensing~\cite{manabe2010licensesevol, debsources-ese-2016,
  vendome2017githublicenses}, on the emergence of open source \emph{license
  variants} and exceptions~\cite{german2015licensevariability,
  vendome2017licexceptions}, as well as the development of industry-strength
tools to automatically detect and classify (FOSS)
licenses~\cite{gobeille2008fossology, german2015licensevariability,
  ombredanne2020sca}.

\paragraph{Contributions and use cases}

We introduce a large-scale dataset of license files collected from more than
150 million public software origins including public Git repositories (from
GitHub and GitLab), FOSS distributions (e.g., Debian), and package manager
repositories (e.g., PyPI, NPM). The dataset is comprised of two parts:
\begin{enumerate}

\item the content of \num{\DataBlobCount} deduplicated license files (or
  \emph{license blobs} in the following) retrieved from \SWH, the largest
  public archive of software source
  code,\footnote{\url{https://archive.softwareheritage.org}, accessed
    2022-01-26} carrying filenames that are commonly used by developers to
  distribute licensing terms to software recipients (e.g., \texttt{COPYING},
  \texttt{LICENSE}, etc.; see \cref{sec:methodology} for details);

\item mined metadata about license files: length measures, detected MIME type,
  contained FOSS license detected using ScanCode~\cite{scancode-toolkit},
  example origin, oldest and total number of public commits in which the
  license file appears.

\end{enumerate}

\noindent
The dataset serves use cases such as: (a) large-scale analyses of open source
licensing, including license popularity, variants, and phylogenetics (how FOSS
licenses evolve and mutate); (b) training supervised and unsupervised machine
learning classifiers for FOSS licenses, which remains an open industry
challenge with most state-of-the-art classifiers still relying on
manually-tuned heuristics; (c) natural language processing (NLP) analyses and
modeling of legal corpora in the semantic domain of software licensing.

\paragraph{Data availability}

The dataset~\cite{stefano_zacchiroli_2022_6379164} is released as open data,
together with a replication package to recreate it from scratch. It is
available for download from Zenodo at
\url{https://doi.org/10.5281/zenodo.6379164} as a \texttt{tar} archive
containing unique license blobs (deduplicated based on SHA1 checksums) in a
sharded directory structure, together with a set of portable CSV files for
derived metadata, cross-referenced to license blobs via SHA1 checksums.

The dataset has been around
informally\footnote{\url{https://annex.softwareheritage.org/public/dataset/license-blobs/},
  accessed 2020-03-23} since 2019 and recently refreshed for the 2021 release
documented in this paper. It has already been used to conduct research
internships in computer forensics (applying TLSH hashing~\cite{oliver2013tlsh}
to measure license distances) and is currently being used to conduct a
large-scale study in open source license phylogenetics.

 \section{Methodology and Reproducibility}
\label{sec:methodology}

\begin{figure}
  \includegraphics[width=\columnwidth]{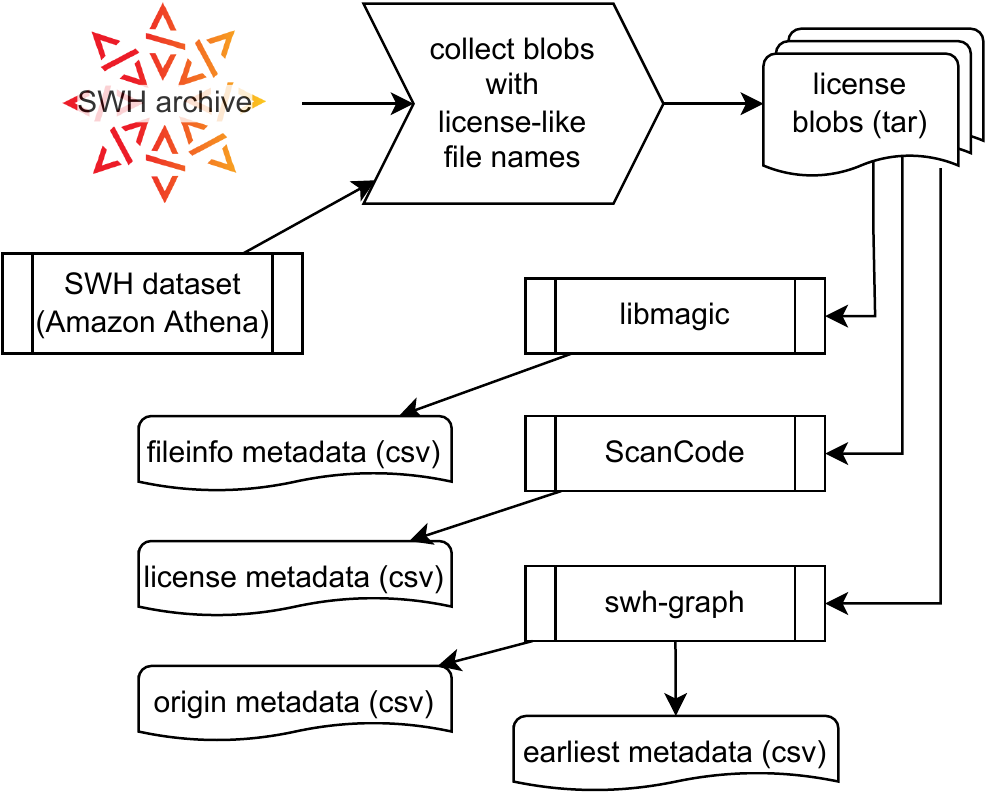}
  \caption{Dataset construction pipeline.}
  \label{fig:methodology}
\end{figure}

\Cref{fig:methodology} depicts the methodology used to assemble the dataset.

\paragraph{License files gathering}

The first step consists in \emph{selecting} all file blobs archived by Software
Heritage and associated to filenames that \emph{are likely} to contain license
texts. To that end we used the Software Heritage graph
dataset~\cite{swh-msr2019-dataset} hosted on Amazon Athena (version 20210323)
to retrieve the SWHID and SHA1 identifiers of all file blobs associated to file
names matching the SQL regular expression:
\verb@^([a-z0-9._-]+\.)?(copying|licen(c|s)(e|ing)|notice@\linebreak \verb@|copyright|disclaimer|authors)(\.[a-z0-9\._-]+)?$@\linebreak (the complete SQL query is available as part of the dataset replication
package, under the \texttt{replication/} directory).

This predicate is quite lax and will end up including files that contain data
other than license texts. This was done on purpose, because while it is trivial
to filter dataset blobs based on filenames using \texttt{fileinfo} metadata
(see \Cref{sec:datamodel}), it is cumbersome to \emph{extend} the dataset
downstream to add all blobs of interest.

We then retrieved all selected blobs from the Software Heritage
archive~\cite{swhipres2017} and archived them in a single \texttt{tar} file.
(This step was conducted in collaboration with the Software Heritage team, but
can be independently reproduced using any archive copy or mirror.)

\paragraph{Metadata mining}

All collected blobs have then been mined to gather various types of metadata
(see \Cref{fig:metadata-schema}, discussed in \Cref{sec:datamodel}). The code
we used for mining is available as part of the dataset replication package.

To detect file MIME types and character encodings, we invoked
libmagic~\cite{file-opengroup} on each blob via the \texttt{python-magic}
Python bindings. For files with MIME type starting with \texttt{text/} and
UTF-8 encoding (or \emph{textual files} in the following for brevity) we
computed line and word counts using custom Python code; for all files we
computed file sizes in bytes.

The likely licenses contained in each blobs have been detected by running the
ScanCode toolkit~\cite{scancode-toolkit} using its Python API. We run ScanCode
with no minimum score threshold---meaning that all detected licenses will be
returned, no matter the tool confidence in the result---and with a timeout of 2
minutes (per blob).

Finally, we used the compressed in-memory graph
representation~\cite{saner-2020-swh-graph} of the Software Heritage archive to
\texttt{origins} and \texttt{earliest} metadata. For origins we used the
\texttt{/randomwalk} API
endpoint\footnote{\url{https://docs.softwareheritage.org/devel/swh-graph/api.html\#get--graph-randomwalk--src--dst},
  accessed 2021-01-25} to traverse the transposed Merkle DAG of the archive and
navigate from each blob to a random origin referencing it.
$\approx$\DataBlobNoOriginPct\% blobs could not be mapped to an origin this way
and lack origin metadata in the dataset.

For earliest commit information we used ad-hoc Java code to navigate the
transposed graph from each blob to all commits referencing it, which were
counted as the number of occurrences of the blob in the archive. Then we
selected the commit with the oldest timestamp among them and extracted its
identifier and Unix time. $\approx$\DataBlobNoEarliestPct\% blobs could not be
mapped to an earliest commit this way and lack earliest metadata in the
dataset.

 \section{Data Model}
\label{sec:datamodel}

\paragraph{License files}

All \num{\DataBlobCount} license blobs are shipped in a single \texttt{tar}
archive file (\texttt{blobs.tar.zst}) compressed with
Zstandard~\cite{zstandard-rfc} and weighting \DataTarSizeApprox. Contained
files are organized in a 2-level-deep sharded directory structure based on the
SHA1 checksum of each file, e.g., {\small
  \texttt{blobs/02/52/0252d93ad297ec183a567ee813ab8c8d61ece655}} for a random
file in the archive. Note hence that license files are \emph{fully
  deduplicated} in the dataset based on SHA1 checksums: each different license
blob will appear exactly once in the archive.

The dataset also includes \texttt{blobs-sample20k.tar.zst}, a smaller archive
containing ``only'' \num{20000} randomly selected license files. It can be used
to conduct trial experiments on a small dataset before attacking the entire
corpus.

\begin{figure}
  \includegraphics[width=\columnwidth]{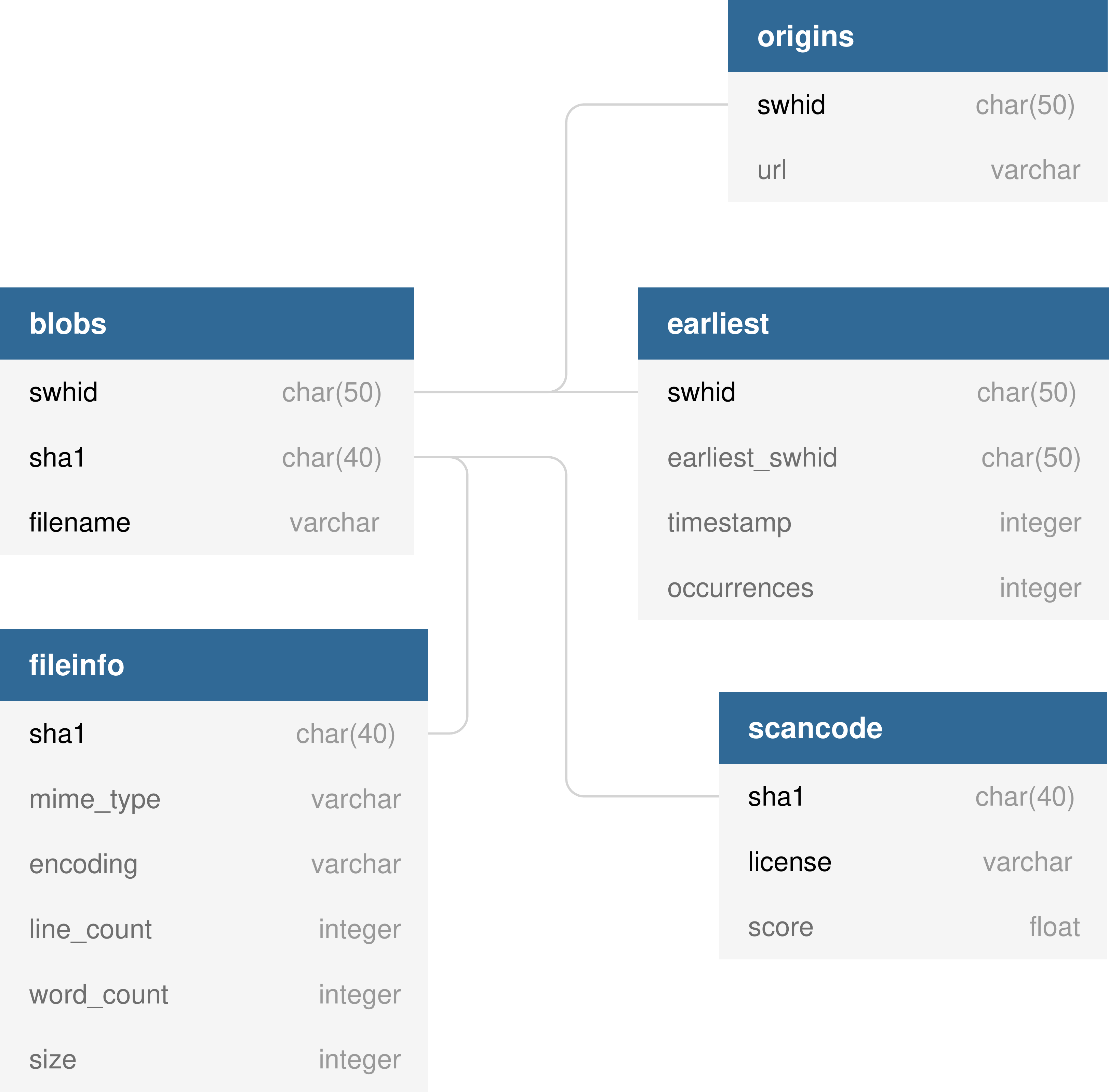}
  \caption{Relational data model for license blob metadata.}
  \label{fig:metadata-schema}
\end{figure}

\paragraph{Metadata}

License file metadata are provided as a set of textual CSV~\cite{csv-rfc}
files, compressed with Zstandard. Each of them corresponds to a table in the
relational model shown in \Cref{fig:metadata-schema}. They can be used as is or
trivially imported into an actual database management system.
Metadata can be cross-referenced to the actual license files (in
\texttt{blobs.tar.zst}) using SHA1 checksums as keys.
Each table captures the metadata described below.

\texttt{\bfseries blobs} (CSV file: \texttt{license-blobs.csv.zst}) is the
master index of all license files (or ``blobs'') in the dataset. The first
column is the Software Heritage persistent identifier
(SWHID)~\cite{swhipres2018} of a blob, e.g., {\footnotesize
  \texttt{swh:1:cnt:94a9ed024d3859793618152ea559a168bbcbb5e2}} for a popular
variant of the GPL version 3 text; the second the SHA1 checksum of the
file. \texttt{filename} is the local name given to this license variant
\emph{in a given context} (e.g., one or more commits in a public Git
repository). This variant of the GPL text is found with 604 different names,
including \texttt{"COPYING"}, \texttt{"LICENSE.GPL3"}, and
\texttt{"a2ps.license"}.  Note that both \texttt{swhid} and \texttt{sha1} are
used by other tables as foreign key targets and that there is no unique primary
key in \texttt{blobs}, due to multiple filenames associated to each license
file.

\texttt{\bfseries fileinfo} (\texttt{blobs-fileinfo.csv.zst}) provides basic
information about license files, cross-referencing them to \texttt{blobs} via
the \texttt{sha1} column. \texttt{mime\_type} and \texttt{encoding} are
respectively the file MIME type and character encoding, as detected by
libmagic~\cite{file-opengroup}. \texttt{size} is the file size in bytes; for
textual files, \texttt{line\_count} and \texttt{word\_count} report file sizes
in lines and (blank-separated) words, respectively.

\texttt{\bfseries scancode} (\texttt{blobs-scancode.csv.zst}) reports about the
license(s) contained in a given file, as detected by the ScanCode
toolkit~\cite{scancode-toolkit, ombredanne2020sca}. Multiple licenses can be
detected within a single file, due to either multiple license texts being
included or to different confidence levels in the answer reported by ScanCode.
For each license file (\texttt{sha1} column), \texttt{license} reports the
license via the associated industry-standard SPDX~\cite{stewart2010spdxspec,
  gandhi2018spdx} identifier (e.g., \texttt{"GPL-3.0-only"}) and \texttt{score}
its confidence level as a float in the $[0,100]$ range (100 being maximum
confidence).

\texttt{\bfseries origins} (\texttt{blobs-origins.csv.zst}) contains
information about where license blobs were found, i.e., which ``projects'' have
distributed them in the past. As each unique license blob can be distributed by
tens of million repositories, only a \emph{single example} of an origin is
given for each license blob via the \texttt{url} field of this table. Obtaining
from \SWH a list of \emph {all} the projects known to ship a given license blob
is possible~\cite{swh-provenance-emse}, but out of scope for this dataset. For
example, the aforementioned variant of the GPL-3 text was found (among others)
in the Git repository at \url{https://github.com/pombreda/Artemis}.

\texttt{\bfseries earliest} (\texttt{blobs-earliest.csv.zst}) provides
historical and popularity information. \texttt{earliest\_swhid} gives the SWHID
of the oldest known public commit that contained the license file, e.g.,
{\footnotesize \texttt{swh:1:rev:088313246501c78ae9d7f08e46aaea45855c5c7e}} for
a variant of the MIT license that includes a Russian copyright notice;
\texttt{timestamp} is the commit timestamp as Unix time. Referenced commit can
be then looked up using the Software Heritage Web UI,\footnote{e.g., said
  Russian MIT variant can be browsed at
  \url{https://archive.softwareheritage.org/swh:1:rev:088313246501c78ae9d7f08e46aaea45855c5c7e}. Accessed
  2021-01-25} API, or filesystem~\cite{swh-fuse}. \texttt{occurrences} reports
the total number of commits known by Software Heritage as containing the
license file; it can be used as a (rough) measure of file popularity.

 \section{Using license (meta)data}
\label{sec:statistics}

We give below some examples of dataset usage, by conducting preliminary
analyses of the license corpus and associated metadata.

\smallskip

Any preliminary analysis of a large textual corpus starts by looking at
\emph{word frequencies}. So let's do that. Iterating on all license blobs to
tokenize, case-normalize, and count words is left as an exercise for the
reader. Assuming a CSV file with $\langle$word, frequency$\rangle$ columns is
produced at the end, the following Python snippet using
Pandas~\cite{mckinney2011pandas} and NLTK~\cite{bird2006nltk} will extract the
top 100 words in the corpus by frequencies, after removal of English stopwords
and single-character tokens.
\begin{lstlisting}[language=Python]
words = pd.read_csv("blobs-wordfreqs.csv") \
  .sort_values(by="frequency", ascending=False)
stop_words = stopwords.words('english') + \
  list(string.digits) + list(string.ascii_lowercase)
interesting_words = words[~words["word")
interesting_words.nlargest(100, columns="frequency")
\end{lstlisting}

\begin{table}
  \caption{Top-10 words in the license corpus by frequency.}
  \label{tab:top-words}
\begin{tabular}{c|r}
    \textbf{Word} & \multicolumn{1}{c}{\textbf{Frequency}} \\
    \hline
    \begin{filecontents*}{words_top1.csv}
word,frequency
software,60539515
license,47336592
copyright,41946018
use,28240621
work,23706422
\end{filecontents*}
\csvreader[
      head to column names,
    ]{words_top1.csv}{}{
      \word & \num{\frequency} \\
    }
  \end{tabular}
  \hspace{1em}
  \begin{tabular}{c|r}
    \textbf{Word} & \multicolumn{1}{c}{\textbf{Frequency}} \\
    \hline
    \begin{filecontents*}{words_top2.csv}
word,frequency
without,22323420
gplv2,21486437
including,20824553
nasl,20746863
notice,19279466
\end{filecontents*}
\csvreader[
      head to column names,
    ]{words_top2.csv}{}{
      \word & \num{\frequency} \\
    }
  \end{tabular}
\end{table}

\Cref{tab:top-words} provides an excerpt of the results, which correspond to
meaningful terms in the semantic domain of open source licensing.

\smallskip

How about \emph{non textual license files}? We can analyze the top detected
MIME types using included \texttt{fileinfo} metadata:
\begin{lstlisting}[language=Python]
fileinfo = pd.read_csv("blobs-fileinfo.csv")
fileinfo["mime_type"].value_counts()
\end{lstlisting}
We omit the results for brevity, but they show that \DataMimeTextPlainPct\% of
the corpus blobs are \texttt{text/plain} and \DataMimeTextPct\% \texttt{text/}
of some kind (including HTML, XML, and LaTeX). Other interesting (small)
classes are rich text formats like RTF as well image files, including PDFs.  We
have manually verified that at least some of these are actually used to
distribute licensing terms; the rest is a small amount of noisy data.

\smallskip

Let's now look at the top \emph{open source licenses detected} in the corpus
files. They are trivial to analyze using the ScanCode metadata included in the
dataset:
\begin{lstlisting}[language=Python]
scancode = pd.read_csv("blobs-scancode.csv")
scancode["license"].value_counts().nlargest(10)
\end{lstlisting}
Note that this is very crude, we are counting all licenses no matter the
associated score; at the same time it is easy to verify (e.g., looking at
\lstinline[language=Python]|scancode["score"].describe()|) that the average
accuracy is very high, with an average of 93 and a 99\% percentile of 100.
More accurate analyses, e.g., of only the licenses detected with score 100
would be trivial to conduct.

\begin{figure}
  \includegraphics[width=\columnwidth,trim=1.5cm 1.5cm 1.5cm 1.5cm,clip]{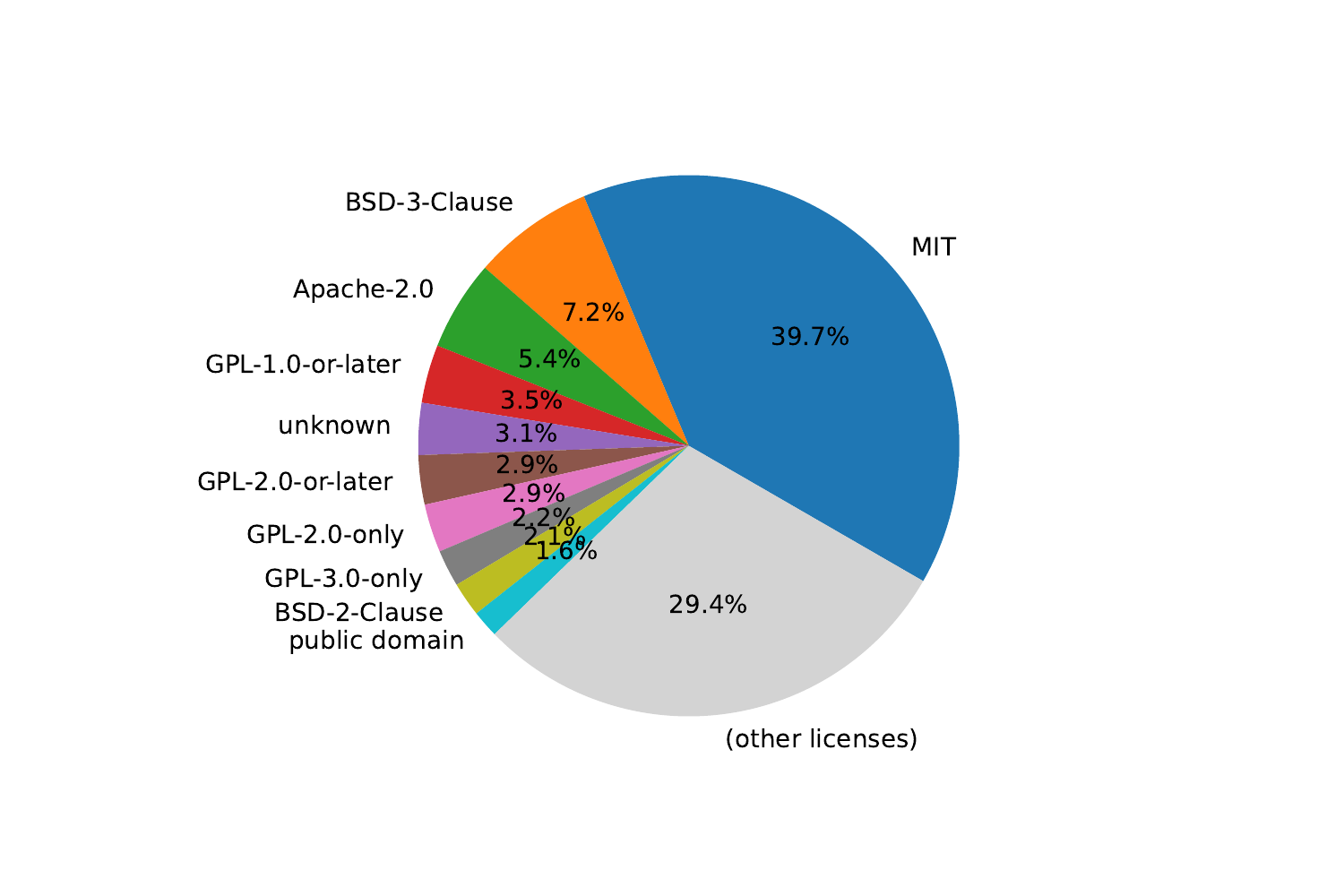}
  \caption{Top licenses in the corpus, as detected by ScanCode.}
  \label{fig:top-licenses}
\end{figure}

Results in \Cref{fig:top-licenses} show that MIT is the most popular open
source license variant in the corpus, followed by 3-clause BSD, and Apache
2. Considering that we are counting \emph{license variants} here, MIT at the
top makes intuitive sense, because its text also includes a copyright notice
which needs to be instantiated by individual authors. Note that this is not a
measure of license \emph{popularity}, but interested researchers can obtain
insights about that by joining these results with \texttt{origin} and
\texttt{earliest} metadata from the dataset.

 \section{Limitations}
\label{sec:discussion}

\paragraph{Internal validity}

Datasets built from large amounts of real-world data tend to be noisy and
contain bogus data (non-license files, in this case). Rather than thoroughly
trying to clean up license blobs incurring the risk of false negatives, we have
decided to \emph{augment} them with extra metadata that enable researcher to
filter data downstream. We have already observed in \Cref{sec:methodology} how
to restrict the filename pattern if so desired. Similarly, researchers can
filter on MIME types (e.g., if only interested in textual files) or on length
metrics (e.g., only keep oneliner files to focus on copyright notices or
machine-readable SPDX tags). Study-specific filtering is also best left to
dataset users and we provide several types of metadata to support it.

The main inconsistency in the dataset come from the incompleteness of
\texttt{origin} and \texttt{earliest} metadata, which are missing for
\DataBlobNoEarliestPct--\DataBlobNoOriginPct\% of the dataset blobs. This is
due to a version misalignment between the Software Heritage archive and the
compressed graph we used for mining these metadata; it could be fixed in the
future when a fresher version will become available. Also, due to the ease of
forging Git timestamps, some earliest commit metadata are bogus having
timestamps set to the UNIX epoch. Both metadata coverage (which remains very
high) and timestamp quality can be improved by cross-referencing license blobs
to external data sources thanks to the persistent identifiers used in the
dataset as keys.

\paragraph{Construct validity}

There is no guarantee that all license blobs in the dataset contain license
texts considered open/free by OSI/FSF (hence the parentheses around ``(open
source)'' in this paper title), only that they come from public code. If
relying on ScanCode as ground truth is acceptable, \texttt{scancode} metadata
in the dataset can be used for filtering. Otherwise the free/open determination
will need to be done independently by dataset users.

Due to selecting license files by filename, \emph{license notices} that
\emph{only appear within} source files are underrepresented in the dataset.
This applies to, e.g., both the recommended GPL notice ``This program is free
software [\ldots] under the terms of the GNU General Public License [\ldots]''
and SPDX tags~\cite{stewart2010spdxspec} like ``SPDX-License-Identifier:
GPL-3.0-or-later'' when they are included only as comments at the beginning of
source files. As the dataset is meant to enable studying license \emph{texts},
rather than notices, this is an acceptable limitation. Also, notices are
included in the dataset when \emph{also} shipped under license-related
filenames. Thoroughly extracting license notices from \SWH and including them
in the dataset is left as future work.

\paragraph{External validity}

By its own nature the dataset provides an incomplete snapshot of reality; as
such we do not claim full generality/representativeness of all existing license
variants. The reality is a moving target, with new license variants constantly
released as public code. The archive we started from is not full-encompassing
either. Still, to the best of our knowledge, this is to date the largest,
publicly available dataset of (open source) license variants. We plan to
mitigate this risk by periodically making available new dataset releases, as we
have done up to now.

 \section{Related Work}
\label{sec:related}

The Software Heritage (SWH) graph dataset~\cite{swh-msr2019-dataset}, which we
used to select license blobs, is a large dataset underpinning the SWH archive.
It stores information analogous to those captured by version control systems
(VCS), minus actual file contents. It can be used in conjunction with the
dataset presented here, joining information via SWH identifiers.

World of Code~\cite{ma2021woc} is a large dataset and analysis infrastructure,
available to researchers to mine public code. It is larger than our initial
data source and can be used in conjunction with this dataset to find additional
origins/occurrences of licenses blobs of interest. Our dataset is smaller, can
be self-hosted, and comes with several relevant metadata precomputed (e.g.,
ScanCode results).

GHTorrent~\cite{GHTorrent} is a dataset of archived GitHub REST API events.  It
contains information about public GitHub projects, but as of today does not
include the license that GitHub detected as the main license of a given
project. (Nor license texts, as source code is out of scope for GHTorrent.)

ScanCode LicenseDB~\cite{scancode-licensedb} is a public database by the
ScanCode authors listing all the licenses they have encountered in the wild
during the constant tuning of their detection heuristics. It includes 1879
different \emph{canonical} license texts which are used as comparison
reference, but does not provide all variants of them as we do with this
dataset; nor it provides associated metadata. Both the Open Source Initiative
and the SPDX project maintain analogous public databases~\cite{osi-licensedb,
  SPDXLicences} covering the canonical texts of, respectively, OSI-approved and
SPDX-named licenses, for about $\approx$500 texts in total.

\smallskip

In summary, this dataset appears to be unique in nature and size, filling an
unattended niche. It can also be used in synergy with preexisting datasets
about FOSS and public code.

 \section{Conclusion}
\label{sec:conclusion}

We have introduced a large-scale dataset of open source license texts. It
consists of \DataBlobCountApproxM{} million unique files archived from public
code and carrying a name related to software licensing terms. Derived
metadata---about file lengths, types, detected open source license in them, and
their provenance---are also included in the dataset and trivial to
cross-reference with the text corpus.

\paragraph{Future extensions}

As future work we intend, on the one hand, to keep the dataset current with the
constant evolution of archived public code, gathering license texts from
additional data sources.  On the other hand we will explore adding to the
metadata precomputed text representations of the entire corpus that are
commonly needed for natural language processing (NLP) and machine learning
analyses, such as word embeddings, latent semantic indexes, and other vectorial
text representations.




\begin{thebibliography}{34}


\ifx \showCODEN    \undefined \def \showCODEN     #1{\unskip}     \fi
\ifx \showDOI      \undefined \def \showDOI       #1{#1}\fi
\ifx \showISBNx    \undefined \def \showISBNx     #1{\unskip}     \fi
\ifx \showISBNxiii \undefined \def \showISBNxiii  #1{\unskip}     \fi
\ifx \showISSN     \undefined \def \showISSN      #1{\unskip}     \fi
\ifx \showLCCN     \undefined \def \showLCCN      #1{\unskip}     \fi
\ifx \shownote     \undefined \def \shownote      #1{#1}          \fi
\ifx \showarticletitle \undefined \def \showarticletitle #1{#1}   \fi
\ifx \showURL      \undefined \def \showURL       {\relax}        \fi
\providecommand\bibfield[2]{#2}
\providecommand\bibinfo[2]{#2}
\providecommand\natexlab[1]{#1}
\providecommand\showeprint[2][]{arXiv:#2}

\bibitem[Allan{\c{c}}on et~al\mbox{.}(2021)]%
        {swh-fuse}
\bibfield{author}{\bibinfo{person}{Thibault Allan{\c{c}}on},
  \bibinfo{person}{Antoine Pietri}, {and} \bibinfo{person}{Stefano
  Zacchiroli}.} \bibinfo{year}{2021}\natexlab{}.
\newblock \showarticletitle{The Software Heritage Filesystem (SwhFS):
  Integrating Source Code Archival with Development}. In
  \bibinfo{booktitle}{\emph{43rd {IEEE/ACM} International Conference on
  Software Engineering: Companion Proceedings, {ICSE} Companion 2021, Madrid,
  Spain, May 25-28, 2021}}. \bibinfo{publisher}{{IEEE}},
  \bibinfo{pages}{45--48}.
\newblock
\urldef\tempurl%
\url{https://doi.org/10.1109/ICSE-Companion52605.2021.00032}
\showDOI{\tempurl}


\bibitem[Bird(2006)]%
        {bird2006nltk}
\bibfield{author}{\bibinfo{person}{Steven Bird}.}
  \bibinfo{year}{2006}\natexlab{}.
\newblock \showarticletitle{{NLTK:} The Natural Language Toolkit}. In
  \bibinfo{booktitle}{\emph{{ACL} 2006, 21st International Conference on
  Computational Linguistics and 44th Annual Meeting of the Association for
  Computational Linguistics, Proceedings of the Conference, Sydney, Australia,
  17-21 July 2006}}, \bibfield{editor}{\bibinfo{person}{Nicoletta Calzolari},
  \bibinfo{person}{Claire Cardie}, {and} \bibinfo{person}{Pierre Isabelle}}
  (Eds.). \bibinfo{publisher}{The Association for Computer Linguistics}.
\newblock
\urldef\tempurl%
\url{https://doi.org/10.3115/1225403.1225421}
\showDOI{\tempurl}


\bibitem[Boldi et~al\mbox{.}(2020)]%
        {saner-2020-swh-graph}
\bibfield{author}{\bibinfo{person}{Paolo Boldi}, \bibinfo{person}{Antoine
  Pietri}, \bibinfo{person}{Sebastiano Vigna}, {and} \bibinfo{person}{Stefano
  Zacchiroli}.} \bibinfo{year}{2020}\natexlab{}.
\newblock \showarticletitle{Ultra-Large-Scale Repository Analysis via Graph
  Compression}. In \bibinfo{booktitle}{\emph{SANER 2020: The 27th IEEE
  International Conference on Software Analysis, Evolution and Reengineering}}.
  \bibinfo{publisher}{IEEE}.
\newblock


\bibitem[Caneill et~al\mbox{.}(2017)]%
        {debsources-ese-2016}
\bibfield{author}{\bibinfo{person}{Matthieu Caneill},
  \bibinfo{person}{Daniel~M. GermÃ¡n}, {and} \bibinfo{person}{Stefano
  Zacchiroli}.} \bibinfo{year}{2017}\natexlab{}.
\newblock \showarticletitle{The Debsources Dataset: Two Decades of Free and
  Open Source Software}.
\newblock \bibinfo{journal}{\emph{Empirical Software Engineering}}
  \bibinfo{volume}{22} (\bibinfo{date}{June} \bibinfo{year}{2017}),
  \bibinfo{pages}{1405--1437}.
\newblock
\showISSN{1382-3256}
\urldef\tempurl%
\url{https://doi.org/10.1007/s10664-016-9461-5}
\showDOI{\tempurl}


\bibitem[Collet(2021)]%
        {zstandard-rfc}
\bibfield{author}{\bibinfo{person}{Yann Collet}.}
  \bibinfo{year}{2021}\natexlab{}.
\newblock \bibinfo{title}{{RFC 8878} - {Zstandard} Compression and the
  ``application/zstd'' Media Type}.
\newblock
\newblock
\urldef\tempurl%
\url{https://datatracker.ietf.org/doc/html/rfc8878}
\showURL{%
\tempurl}
\newblock
\shownote{Accessed 2022-01-24}.


\bibitem[Di~Cosmo et~al\mbox{.}(2018)]%
        {swhipres2018}
\bibfield{author}{\bibinfo{person}{Roberto Di~Cosmo}, \bibinfo{person}{Morane
  Gruenpeter}, {and} \bibinfo{person}{Stefano Zacchiroli}.}
  \bibinfo{year}{2018}\natexlab{}.
\newblock \showarticletitle{Identifiers for Digital Objects: the Case of
  Software Source Code Preservation}. In \bibinfo{booktitle}{\emph{Proceedings
  of the 15th International Conference on Digital Preservation, iPRES 2018,
  Boston, USA}}.
\newblock
\urldef\tempurl%
\url{https://doi.org/10.17605/OSF.IO/KDE56}
\showDOI{\tempurl}


\bibitem[Di~Cosmo and Zacchiroli(2017)]%
        {swhipres2017}
\bibfield{author}{\bibinfo{person}{Roberto Di~Cosmo} {and}
  \bibinfo{person}{Stefano Zacchiroli}.} \bibinfo{year}{2017}\natexlab{}.
\newblock \showarticletitle{{Software Heritage}: Why and How to Preserve
  Software Source Code}. In \bibinfo{booktitle}{\emph{Proceedings of the 14th
  International Conference on Digital Preservation, iPRES 2017}}.
\newblock
\urldef\tempurl%
\url{https://hal.archives-ouvertes.fr/hal-01590958/}
\showURL{%
\tempurl}


\bibitem[Gandhi et~al\mbox{.}(2018)]%
        {gandhi2018spdx}
\bibfield{author}{\bibinfo{person}{Robin~A. Gandhi}, \bibinfo{person}{Matt
  Germonprez}, {and} \bibinfo{person}{Georg J.~P. Link}.}
  \bibinfo{year}{2018}\natexlab{}.
\newblock \showarticletitle{Open Data Standards for Open Source Software Risk
  Management Routines: An Examination of {SPDX}}. In
  \bibinfo{booktitle}{\emph{Proceedings of the 2018 {ACM} Conference on
  Supporting Groupwork, {GROUP} 2018, Sanibel Island, FL, USA, January 07 - 10,
  2018}}, \bibfield{editor}{\bibinfo{person}{Andrea Forte},
  \bibinfo{person}{Michael Prilla}, \bibinfo{person}{Adriana~S. Vivacqua},
  \bibinfo{person}{Claudia M{\"{u}}ller}, {and} \bibinfo{person}{Lionel
  P.~Robert Jr.}} (Eds.). \bibinfo{publisher}{{ACM}},
  \bibinfo{pages}{219--229}.
\newblock
\urldef\tempurl%
\url{https://doi.org/10.1145/3148330.3148333}
\showDOI{\tempurl}


\bibitem[Germ{\'{a}}n and Penta(2012)]%
        {german2012liccompatibility}
\bibfield{author}{\bibinfo{person}{Daniel~M. Germ{\'{a}}n} {and}
  \bibinfo{person}{Massimiliano~Di Penta}.} \bibinfo{year}{2012}\natexlab{}.
\newblock \showarticletitle{A Method for Open Source License Compliance of Java
  Applications}.
\newblock \bibinfo{journal}{\emph{{IEEE} Softw.}} \bibinfo{volume}{29},
  \bibinfo{number}{3} (\bibinfo{year}{2012}), \bibinfo{pages}{58--63}.
\newblock
\urldef\tempurl%
\url{https://doi.org/10.1109/MS.2012.50}
\showDOI{\tempurl}


\bibitem[Gobeille(2008)]%
        {gobeille2008fossology}
\bibfield{author}{\bibinfo{person}{Robert Gobeille}.}
  \bibinfo{year}{2008}\natexlab{}.
\newblock \showarticletitle{The FOSSology project}. In
  \bibinfo{booktitle}{\emph{Proceedings of the 2008 International Working
  Conference on Mining Software Repositories, {MSR} 2008 (Co-located with
  ICSE), Leipzig, Germany, May 10-11, 2008, Proceedings}},
  \bibfield{editor}{\bibinfo{person}{Ahmed~E. Hassan}, \bibinfo{person}{Michele
  Lanza}, {and} \bibinfo{person}{Michael~W. Godfrey}} (Eds.).
  \bibinfo{publisher}{{ACM}}, \bibinfo{pages}{47--50}.
\newblock
\urldef\tempurl%
\url{https://doi.org/10.1145/1370750.1370763}
\showDOI{\tempurl}


\bibitem[Gousios and Spinellis(2012)]%
        {GHTorrent}
\bibfield{author}{\bibinfo{person}{Georgios Gousios} {and}
  \bibinfo{person}{Diomidis Spinellis}.} \bibinfo{year}{2012}\natexlab{}.
\newblock \showarticletitle{GHTorrent: Github's data from a firehose}. In
  \bibinfo{booktitle}{\emph{9th {IEEE} Working Conference of Mining Software
  Repositories, {MSR}}}, \bibfield{editor}{\bibinfo{person}{Michele Lanza},
  \bibinfo{person}{Massimiliano~Di Penta}, {and} \bibinfo{person}{Tao Xie}}
  (Eds.). \bibinfo{publisher}{{IEEE} Computer Society},
  \bibinfo{pages}{12--21}.
\newblock
\showISBNx{978-1-4673-1761-0}
\urldef\tempurl%
\url{https://doi.org/10.1109/MSR.2012.6224294}
\showDOI{\tempurl}


\bibitem[Harutyunyan(2020)]%
        {harutyunyan2020osssupplychain}
\bibfield{author}{\bibinfo{person}{Nikolay Harutyunyan}.}
  \bibinfo{year}{2020}\natexlab{}.
\newblock \showarticletitle{Managing Your Open Source Supply Chain-Why and
  How?}
\newblock \bibinfo{journal}{\emph{Computer}} \bibinfo{volume}{53},
  \bibinfo{number}{6} (\bibinfo{year}{2020}), \bibinfo{pages}{77--81}.
\newblock
\urldef\tempurl%
\url{https://doi.org/10.1109/MC.2020.2983530}
\showDOI{\tempurl}


\bibitem[Lindberg(2008)]%
        {lindberg2008osslicensing}
\bibfield{author}{\bibinfo{person}{Van Lindberg}.}
  \bibinfo{year}{2008}\natexlab{}.
\newblock \bibinfo{booktitle}{\emph{Intellectual property and open source: a
  practical guide to protecting code}}.
\newblock \bibinfo{publisher}{O'Reilly Media, Inc.}
\newblock


\bibitem[Ma et~al\mbox{.}(2021)]%
        {ma2021woc}
\bibfield{author}{\bibinfo{person}{Yuxing Ma}, \bibinfo{person}{Tapajit Dey},
  \bibinfo{person}{Chris Bogart}, \bibinfo{person}{Sadika Amreen},
  \bibinfo{person}{Marat Valiev}, \bibinfo{person}{Adam Tutko},
  \bibinfo{person}{David Kennard}, \bibinfo{person}{Russell Zaretzki}, {and}
  \bibinfo{person}{Audris Mockus}.} \bibinfo{year}{2021}\natexlab{}.
\newblock \showarticletitle{World of code: enabling a research workflow for
  mining and analyzing the universe of open source {VCS} data}.
\newblock \bibinfo{journal}{\emph{Empir. Softw. Eng.}} \bibinfo{volume}{26},
  \bibinfo{number}{2} (\bibinfo{year}{2021}), \bibinfo{pages}{22}.
\newblock
\urldef\tempurl%
\url{https://doi.org/10.1007/s10664-020-09905-9}
\showDOI{\tempurl}


\bibitem[Manabe et~al\mbox{.}(2010)]%
        {manabe2010licensesevol}
\bibfield{author}{\bibinfo{person}{Yuki Manabe}, \bibinfo{person}{Yasuhiro
  Hayase}, {and} \bibinfo{person}{Katsuro Inoue}.}
  \bibinfo{year}{2010}\natexlab{}.
\newblock \showarticletitle{Evolutional analysis of licenses in {FOSS}}. In
  \bibinfo{booktitle}{\emph{Proceedings of the Joint {ERCIM} Workshop on
  Software Evolution {(EVOL)} and International Workshop on Principles of
  Software Evolution (IWPSE), Antwerp, Belgium, September 20-21, 2010}},
  \bibfield{editor}{\bibinfo{person}{Andrea Capiluppi},
  \bibinfo{person}{Anthony Cleve}, {and} \bibinfo{person}{Naouel Moha}} (Eds.).
  \bibinfo{publisher}{{ACM}}, \bibinfo{pages}{83--87}.
\newblock
\urldef\tempurl%
\url{https://doi.org/10.1145/1862372.1862391}
\showDOI{\tempurl}


\bibitem[Maryka et~al\mbox{.}(2015)]%
        {german2015licensevariability}
\bibfield{author}{\bibinfo{person}{Trevor Maryka}, \bibinfo{person}{Daniel~M.
  Germ{\'{a}}n}, {and} \bibinfo{person}{Germ{\'{a}}n Poo{-}Caama{\~{n}}o}.}
  \bibinfo{year}{2015}\natexlab{}.
\newblock \showarticletitle{On the Variability of the {BSD} and {MIT}
  Licenses}. In \bibinfo{booktitle}{\emph{Open Source Systems: Adoption and
  Impact - 11th {IFIP} {WG} 2.13 International Conference, {OSS} 2015,
  Florence, Italy, May 16-17, 2015, Proceedings}}
  \emph{(\bibinfo{series}{{IFIP} Advances in Information and Communication
  Technology}, Vol.~\bibinfo{volume}{451})},
  \bibfield{editor}{\bibinfo{person}{Ernesto Damiani}, \bibinfo{person}{Fulvio
  Frati}, \bibinfo{person}{Dirk Riehle}, {and} \bibinfo{person}{Anthony~I.
  Wasserman}} (Eds.). \bibinfo{publisher}{Springer}, \bibinfo{pages}{146--156}.
\newblock
\urldef\tempurl%
\url{https://doi.org/10.1007/978-3-319-17837-0\_14}
\showDOI{\tempurl}


\bibitem[McKinney et~al\mbox{.}(2011)]%
        {mckinney2011pandas}
\bibfield{author}{\bibinfo{person}{Wes McKinney} {et~al\mbox{.}}}
  \bibinfo{year}{2011}\natexlab{}.
\newblock \showarticletitle{pandas: a foundational Python library for data
  analysis and statistics}.
\newblock \bibinfo{journal}{\emph{Python for high performance and scientific
  computing}} \bibinfo{volume}{14}, \bibinfo{number}{9} (\bibinfo{year}{2011}),
  \bibinfo{pages}{1--9}.
\newblock


\bibitem[{nexB}(2022a)]%
        {scancode-toolkit}
\bibfield{author}{\bibinfo{person}{{nexB}}.} \bibinfo{year}{2022}\natexlab{a}.
\newblock \bibinfo{title}{{ScanCode}}.
\newblock
\newblock
\urldef\tempurl%
\url{https://www.aboutcode.org/projects/scancode.html}
\showURL{%
\tempurl}
\newblock
\shownote{Accessed 2022-01-25}.


\bibitem[{nexB}(2022b)]%
        {scancode-licensedb}
\bibfield{author}{\bibinfo{person}{{nexB}}.} \bibinfo{year}{2022}\natexlab{b}.
\newblock \bibinfo{title}{{ScanCode LicenseDB}}.
\newblock
\newblock
\urldef\tempurl%
\url{https://scancode-licensedb.aboutcode.org/}
\showURL{%
\tempurl}
\newblock
\shownote{Accessed 2022-01-26}.


\bibitem[Oliver et~al\mbox{.}(2013)]%
        {oliver2013tlsh}
\bibfield{author}{\bibinfo{person}{Jonathan Oliver}, \bibinfo{person}{Chun
  Cheng}, {and} \bibinfo{person}{Yanggui Chen}.}
  \bibinfo{year}{2013}\natexlab{}.
\newblock \showarticletitle{{TLSH}: a locality sensitive hash}. In
  \bibinfo{booktitle}{\emph{2013 Fourth Cybercrime and Trustworthy Computing
  Workshop}}. IEEE, \bibinfo{pages}{7--13}.
\newblock


\bibitem[Ombredanne(2020)]%
        {ombredanne2020sca}
\bibfield{author}{\bibinfo{person}{Philippe Ombredanne}.}
  \bibinfo{year}{2020}\natexlab{}.
\newblock \showarticletitle{Free and Open Source Software License Compliance:
  Tools for Software Composition Analysis}.
\newblock \bibinfo{journal}{\emph{Computer}} \bibinfo{volume}{53},
  \bibinfo{number}{10} (\bibinfo{year}{2020}), \bibinfo{pages}{105--109}.
\newblock
\urldef\tempurl%
\url{https://doi.org/10.1109/MC.2020.3011082}
\showDOI{\tempurl}


\bibitem[{Open Source Initiative}(2022)]%
        {osi-licensedb}
\bibfield{author}{\bibinfo{person}{{Open Source Initiative}}.}
  \bibinfo{year}{2022}\natexlab{}.
\newblock \bibinfo{title}{Machine readable {OSI} license information}.
\newblock
\newblock
\urldef\tempurl%
\url{https://github.com/OpenSourceOrg/licenses/}
\showURL{%
\tempurl}
\newblock
\shownote{Accessed 2022-01-26}.


\bibitem[Phipps and Zacchiroli(2020)]%
        {phipps2020continuouscompliance}
\bibfield{author}{\bibinfo{person}{Simon Phipps} {and} \bibinfo{person}{Stefano
  Zacchiroli}.} \bibinfo{year}{2020}\natexlab{}.
\newblock \showarticletitle{Continuous Open Source License Compliance}.
\newblock \bibinfo{journal}{\emph{Computer}} \bibinfo{volume}{53},
  \bibinfo{number}{12} (\bibinfo{year}{2020}), \bibinfo{pages}{115--119}.
\newblock
\urldef\tempurl%
\url{https://doi.org/10.1109/MC.2020.3024403}
\showDOI{\tempurl}


\bibitem[Pietri et~al\mbox{.}(2019)]%
        {swh-msr2019-dataset}
\bibfield{author}{\bibinfo{person}{Antoine Pietri}, \bibinfo{person}{Diomidis
  Spinellis}, {and} \bibinfo{person}{Stefano Zacchiroli}.}
  \bibinfo{year}{2019}\natexlab{}.
\newblock \showarticletitle{The {S}oftware {H}eritage graph dataset: public
  software development under one roof}. In
  \bibinfo{booktitle}{\emph{Proceedings of the 16th International Conference on
  Mining Software Repositories, {MSR} 2019, 26-27 May 2019, Montreal,
  Canada.}}, \bibfield{editor}{\bibinfo{person}{Margaret{-}Anne~D. Storey},
  \bibinfo{person}{Bram Adams}, {and} \bibinfo{person}{Sonia Haiduc}} (Eds.).
  \bibinfo{publisher}{{IEEE} / {ACM}}, \bibinfo{pages}{138--142}.
\newblock
\showISBNx{978-1-7281-3412-3}
\urldef\tempurl%
\url{https://dl.acm.org/citation.cfm?id=3341907}
\showURL{%
\tempurl}


\bibitem[Rosen(2005)]%
        {rosen2005osslicensing}
\bibfield{author}{\bibinfo{person}{Lawrence Rosen}.}
  \bibinfo{year}{2005}\natexlab{}.
\newblock \bibinfo{booktitle}{\emph{Open source licensing}}.
  Vol.~\bibinfo{volume}{692}.
\newblock \bibinfo{publisher}{Prentice Hall}.
\newblock


\bibitem[Rousseau et~al\mbox{.}(2020)]%
        {swh-provenance-emse}
\bibfield{author}{\bibinfo{person}{Guillaume Rousseau},
  \bibinfo{person}{Roberto Di~Cosmo}, {and} \bibinfo{person}{Stefano
  Zacchiroli}.} \bibinfo{year}{2020}\natexlab{}.
\newblock \showarticletitle{Software Provenance Tracking at the Scale of Public
  Source Code}.
\newblock \bibinfo{journal}{\emph{Empirical Software Engineering}}
  \bibinfo{volume}{25}, \bibinfo{number}{4} (\bibinfo{year}{2020}),
  \bibinfo{pages}{2930--2959}.
\newblock
\showISSN{1382-3256}
\urldef\tempurl%
\url{https://doi.org/10.1007/s10664-020-09828-5}
\showDOI{\tempurl}


\bibitem[Shafranovich(2005)]%
        {csv-rfc}
\bibfield{author}{\bibinfo{person}{Yakov Shafranovich}.}
  \bibinfo{year}{2005}\natexlab{}.
\newblock \bibinfo{title}{{RFC 4180} - Common Format and {MIME} Type for
  Comma-Separated Values ({CSV}) Files}.
\newblock
\newblock
\urldef\tempurl%
\url{https://datatracker.ietf.org/doc/html/rfc4180}
\showURL{%
\tempurl}
\newblock
\shownote{Accessed 2022-01-24}.


\bibitem[{SPDX Workgroup}(2019)]%
        {SPDXLicences}
\bibfield{author}{\bibinfo{person}{{SPDX Workgroup}}.}
  \bibinfo{year}{2019}\natexlab{}.
\newblock \bibinfo{title}{Software Package Data Exchange Licence List}.
\newblock
\newblock
\urldef\tempurl%
\url{https://spdx.org/license-list}
\showURL{%
\tempurl}
\newblock
\shownote{\url{https://spdx.org/license-list}, retrieved 30 March 2020}.


\bibitem[Stewart et~al\mbox{.}(2010)]%
        {stewart2010spdxspec}
\bibfield{author}{\bibinfo{person}{Kate Stewart}, \bibinfo{person}{Phil
  Odence}, {and} \bibinfo{person}{Esteban Rockett}.}
  \bibinfo{year}{2010}\natexlab{}.
\newblock \showarticletitle{Software package data exchange ({SPDX})
  specification}.
\newblock \bibinfo{journal}{\emph{IFOSS L. Rev.}}  \bibinfo{volume}{2}
  (\bibinfo{year}{2010}), \bibinfo{pages}{191}.
\newblock


\bibitem[Synopsis(2020)]%
        {synopsis2020ossra}
\bibfield{author}{\bibinfo{person}{Synopsis}.} \bibinfo{year}{2020}\natexlab{}.
\newblock \bibinfo{booktitle}{\emph{2020 Open Source Security and Risk Analysis
  Report {(OSSRA)}}}.
\newblock \bibinfo{type}{{T}echnical {R}eport}.
  \bibinfo{institution}{Synopsis}.
\newblock
\urldef\tempurl%
\url{https://www.synopsys.com/content/dam/synopsys/sig-assets/reports/2020-ossra-report.pdf}
\showURL{%
\tempurl}
\newblock
\shownote{Accessed 2020-04-15}.


\bibitem[{The Open Group}(2018)]%
        {file-opengroup}
\bibfield{author}{\bibinfo{person}{{The Open Group}}.}
  \bibinfo{year}{2018}\natexlab{}.
\newblock \bibinfo{title}{file: determine file type}.
\newblock
\newblock
\urldef\tempurl%
\url{https://pubs.opengroup.org/onlinepubs/9699919799/utilities/file.html}
\showURL{%
\tempurl}
\newblock
\shownote{Accessed 2022-01-25}.


\bibitem[Vendome et~al\mbox{.}(2017a)]%
        {vendome2017githublicenses}
\bibfield{author}{\bibinfo{person}{Christopher Vendome},
  \bibinfo{person}{Gabriele Bavota}, \bibinfo{person}{Massimiliano~Di Penta},
  \bibinfo{person}{Mario~Linares V{\'{a}}squez}, \bibinfo{person}{Daniel~M.
  Germ{\'{a}}n}, {and} \bibinfo{person}{Denys Poshyvanyk}.}
  \bibinfo{year}{2017}\natexlab{a}.
\newblock \showarticletitle{License usage and changes: a large-scale study on
  gitHub}.
\newblock \bibinfo{journal}{\emph{Empir. Softw. Eng.}} \bibinfo{volume}{22},
  \bibinfo{number}{3} (\bibinfo{year}{2017}), \bibinfo{pages}{1537--1577}.
\newblock
\urldef\tempurl%
\url{https://doi.org/10.1007/s10664-016-9438-4}
\showDOI{\tempurl}


\bibitem[Vendome et~al\mbox{.}(2017b)]%
        {vendome2017licexceptions}
\bibfield{author}{\bibinfo{person}{Christopher Vendome},
  \bibinfo{person}{Mario~Linares V{\'{a}}squez}, \bibinfo{person}{Gabriele
  Bavota}, \bibinfo{person}{Massimiliano~Di Penta}, \bibinfo{person}{Daniel~M.
  Germ{\'{a}}n}, {and} \bibinfo{person}{Denys Poshyvanyk}.}
  \bibinfo{year}{2017}\natexlab{b}.
\newblock \showarticletitle{Machine learning-based detection of open source
  license exceptions}. In \bibinfo{booktitle}{\emph{Proceedings of the 39th
  International Conference on Software Engineering, {ICSE} 2017, Buenos Aires,
  Argentina, May 20-28, 2017}},
  \bibfield{editor}{\bibinfo{person}{Sebasti{\'{a}}n Uchitel},
  \bibinfo{person}{Alessandro Orso}, {and} \bibinfo{person}{Martin~P.
  Robillard}} (Eds.). \bibinfo{publisher}{{IEEE} / {ACM}},
  \bibinfo{pages}{118--129}.
\newblock
\urldef\tempurl%
\url{https://doi.org/10.1109/ICSE.2017.19}
\showDOI{\tempurl}


\bibitem[Zacchiroli(2022)]%
        {stefano_zacchiroli_2022_6379164}
\bibfield{author}{\bibinfo{person}{Stefano Zacchiroli}.}
  \bibinfo{year}{2022}\natexlab{}.
\newblock \bibinfo{booktitle}{\emph{A Large-scale Dataset of (Open Source)
  License Text Variants}}.
\newblock
\urldef\tempurl%
\url{https://doi.org/10.5281/zenodo.6379164}
\showDOI{\tempurl}


\end{thebibliography}
\end{document}